\documentclass[showpacs,preprintnumbers,amsmath,amssymb]{revtex4}
\usepackage{amsmath,amssymb,graphics,epsfig,subfigure}
\usepackage{color}
\usepackage[normalem]{ulem}

\begin{document}
\renewcommand{\baselinestretch}{1.3}

\title{Shadows of Kerr-like black holes in a modified gravity theory}

\author{Hui-Min Wang$^{1}$ \footnote{wanghm17@lzu.edu.cn},
        Yu-Meng Xu$^{1}$ \footnote{xuym17@lzu.edu.cn}, Shao-Wen Wei$^{1,2}$ \footnote{weishw@lzu.edu.cn, corresponding author}}

\affiliation{$^{1}$ Institute of Theoretical Physics $\&$ Research Center of Gravitation, Lanzhou University, Lanzhou 730000, People's Republic of China\\
$^{2}$ Department of Physics and Astronomy, University of Waterloo, Waterloo, Ontario, Canada, N2L 3G1}

\begin{abstract}
In this paper, the shadows cast by non-rotating and rotating modified gravity black holes are investigated. In addition to the black hole spin parameter $a$ and the inclination angle $\theta$ of observer, another parameter $\alpha$ measuring the deviation of gravitational constant from the Newton one is also found to affect the shape of the black hole shadow. The result shows that, for fixed values of $a/M$ and $\theta$, the size and perimeter of the shadows cast by the non-rotating and rotating black holes significantly increase with the parameter $\alpha$, while the distortions decrease with $\alpha$. Moreover, the energy emission rate of the black hole in high energy case is also investigated, and the result shows that the peak of the emission rate decreases with the parameter $\alpha$.
\end{abstract}



\maketitle

\section{Introduction}
\label{secIntroduction}

Very recently, it was very exciting that the gravitational waves have been directly observed by LIGO and Virgo from binary black hole and neutron star systems \cite{Abbott}. These events opened a new window for the observational astronomy. As expected, the nature of black holes will be well determined in the near future when more merging events are observed and the signal-to-noise ratio is improved. These will also provide us a possible way to distinguish black holes in different gravity theories  \cite{KonoplyaZh,Bambij} .

There are other ways to explore the nature of black holes, such as strong black hole lensing and shadow. In particular, these effects will be greatly improved by the black hole mass. More interestingly, it is generally believed that there exists a supermassive black hole in the center of each galaxy. The mass of the corresponding black hole is so huge that the strong gravity effects may be observable. For example, the Event Horizon Telescope with high enough angular resolution is able to observe the shadow cast by the supermassive black hole located in the center of our galaxy \cite{Fish} with the new imaging techniques. Therefore, it is extremely desirable to investigate the black hole shadow.

Near the black hole shadow, the strong gravitational lensing may be quite obvious. In the ideal situation, a large Einstein ring and two infinite series of concentric relativistic Einstein rings will appear near the shadow of a non-rotating black hole, while only one or two bright images for the rotating case. Compared with the one-dimensional lensing images, black hole shadows are two-dimensional dark zones, which implies that shadows are much more easily to be observed.

Black hole shadows are formed by the null geodesics in the strong gravity region. In general, these photons with large angular momentum coming from infinity will be bounded back to infinity by the gravitational potential of the black hole. While these with small angular momentum will fall into the black hole, which leads to a dark zone for the observer located at infinity. Among these two cases, the photons with critical angular momentum will round the black hole one loop by one loop, which forms the boundary of the shadow. For the non-rotating Schwarzschild black hole, its shadow was first studied by Synge and Luminet \cite{Synge,Luminet}. The shadow cast by rotating Kerr black hole was first investigated by Bardeen \cite{Bardeen}, and was systematically introduced in Ref. \cite{Chandrasekhar}. These results show that the non-rotating black hole has a perfect circular shadow, while the shape of the rotating black hole shadow is elongated due to the dragging effect.

Importantly, astronomical observables are the key to match the theoretical study and astronomical observation. Considering that the shape of the black hole shadow is uniquely determined by its boundary, observables can be constructed by these characteristic points on the boundary. Hioki and Maeda \cite{Hioki2} proposed several new observables to describe the shadows cast by the Kerr black holes and Kerr naked singularities. Using these observables, the spin of the black hole and the inclination angle of the observer will be well determined. There are other observables constructed by different research groups \cite{Johannsen,Ghasemi,Abdujabbarov}. In particular, the authors in Ref. \cite{Abdujabbarov} showed a new formalism to describe the shapes of the black hole shadows. Based on it, several distortion parameters of the shadow were introduced. This method is independent of the location of the center of the shadow. These studies were also carried out for other black hole or wormhole backgrounds \cite{Falcke,Takahashi,Hioki,Bambi,Kraniotis,Bozzagrg,Schee,Amarilla,Stuchlik,AmarillaEiroa,YumotoNitta,Amarilla13,Nedkova,Wei,Tsukamoto,Atamurotov,Atamurotov2,Mann,WWei,FAtamurotov,FAtamurotov2,Abdujabbarov3,Amir,AmirAhmedov,Songbai,Balendra,Tsukamoto2,Jiliang,Mingzhi,Tsvetan,Rajibul,hou,Cunha,Cunha2,Cunha3,Ali1,Mann3}.

On the other hand, a new interesting modified gravity (MOG), the scalar-tensor-vector gravity theory proposed by Moffat, can be treated as an alternative to general relativity (GR) when one deals with the galaxy rotation curves and galaxy clusters without introducing the dark matter \cite{Moffat0,Rahvar,MoffatR,Toth}. So it is of great interest to investigate the properties of the black hole in this gravity theory. Many studies demonstrated that there exist distinct differences between this MOG and GR \cite{Moffat3,Moffat4,Mureika,Lee,Sharif,Hussain,Perez,Sheoran,Moffat5,Manfredi,WeiLiu,Ali2}. In Refs. \cite{Moffat4,Gguo}, the authors gave a preliminary image of the shadow cast by the MOG black hole. And recently, the superradiance in the MOG was also studied in Ref. \cite{Moffat6}. In this paper, we aim to investigate the observables of the MOG black hole shadows. These results will provide us with the detailed property of the MOG theory, and a possible way to distinguish the MOG from GR.

This paper is organized as follows. In Sec. \ref{Classification}, we briefly summarize the field equations and the MOG black hole metric. The geodesic equations and orbital equations of photons are given in Sec. \ref{phorb}. Further, in Sec. \ref{sshadow}, we examine the shadows of black holes, mainly including the apparent shapes and the distortions of the shadow. Moreover, the energy emission rate is also investigated. And finally, we give a summary and discussion for this article in Sec. \ref{discussions}.

\section{Field equation and MOG black hole}
\label{Classification}

The action of the modified gravity we focus on is \cite{Moffat0}
\begin{eqnarray}
 S=S_{\texttt{GR}}+S_{\phi}+S_{\rm s}+S_{\texttt{M}},
\end{eqnarray}
with the parts $S_{\texttt{GR}}$, $S_{\phi}$, and $S_{\rm s}$ given by
\begin{eqnarray}
 S_{\texttt{GR}}&=&\frac{1}{16\pi}\int d^{4}x\sqrt{-g}\frac{R}{G},\\
 S_{\phi}&=&\int d^{4}x\sqrt{-g}\,\Big(-\frac{1}{4}B^{\mu\nu}B_{\mu\nu}
     +\frac{1}{2}\mu^{2}\phi^{\mu}\phi_{\mu}\Big),\\
S_{\rm S}&=&\int
d^4x\sqrt{-g}\bigg[\frac{1}{G^3}\Big(\frac{1}{2}g^{\mu\nu}\nabla_\mu
G\nabla_\nu G-V(G)\Big)\nonumber\\
&+&\frac{1}{\mu^2G}\Big(\frac{1}{2}g^{\mu\nu}\nabla_\mu\mu\nabla_\nu\mu
-V(\mu)\Big)\bigg].
\end{eqnarray}
Here $\phi^{\mu}$ is a Proca type massive vector field of mass $\mu$. Potentials $V(G)$ and $V(\mu)$ are, respectively, correspond to scalar fields $G(x)$ and $\mu(x)$. $S_{\texttt{M}}$ denotes the matter action. The tensor field $B_{\mu\nu}=\partial_{\mu}\phi_{\nu}-\partial_{\nu}\phi_{\mu}$ satisfies the following equations
\begin{eqnarray}
 &&\nabla_{\nu}B^{\mu\nu}=0,\\
 &&\nabla_{\sigma}B_{\mu\nu}+\nabla_{\mu}B_{\nu\sigma}+\nabla_{\nu}B_{\sigma\mu}=0.
\end{eqnarray}
Since the effect of the mass $\mu$ of the vector field displays at kiloparsec scales from the source, it can be neglected for a black hole solution.  On the other hand, one can also treat $G$ as a constant independent of the spacetime coordinates. Moreover, considering a vacuum solution, the action will be simplified to
\begin{eqnarray}
 S=\int d^{4}x\sqrt{-g}\left(\frac{R}{16\pi G}-\frac{1}{4}B^{\mu\nu}B_{\mu\nu}\right).
\end{eqnarray}
The field equation corresponding to this action reads
\begin{eqnarray}
 G_{\mu\nu}=-8\pi G T_{\phi\mu\nu},\label{field}
\end{eqnarray}
with the energy momentum tensor of the vector field given by
\begin{eqnarray}
 T_{\phi\mu\nu}&=&-\frac{1}{4\pi}\left(B_{\mu}^{\;\sigma}B_{\nu\sigma}
      -\frac{1}{4}g_{\mu\nu}B^{\sigma\beta}B_{\sigma\beta}\right).
\end{eqnarray}
The parameter $G$ has a relation with the Newton's gravitational constant $G=G_{\rm N}(1+\alpha)$ with $\alpha$ being a dimensionless parameter. For $\alpha=0$, it will back to GR. Thus we can treat $\alpha$ as a deviation parameter of the MOG from GR.

Solving these field equations, the rotating Kerr-MOG black hole can be obtained, which in Boyer-Lindquist coordinates has the form \cite{Moffat3}
\begin{equation}
ds^2=-\frac{\Delta-a^2\sin^2\theta}{\rho^2}dt^2+\sin^2\theta\biggl[\frac{(r^2+a^2)^2-\Delta a^2\sin^2\theta}{\rho^2}\biggr]d\phi^2
$$ $$
-2a\sin^2\theta\biggl(\frac{r^2+a^2-\Delta}{\rho^2}\biggr)dtd\phi+\frac{\rho^2}{\Delta}dr^2+\rho^2d\theta^2,
\end{equation}
where
\begin{equation}
\Delta=r^2-2GMr+a^2+\alpha G_{\rm N}GM^2,\quad \rho^2=r^2+a^2\cos^2\theta.
\end{equation}
The Newton mass $M$ and the ADM mass $M_{\rm ADM}$ are related with \cite{Sheoran}
\begin{equation}
 M_{\rm ADM}=(1+\alpha)M.
\end{equation}
Since the shadow is mainly caused by the Newton mass, we will adopt the Newton mass as the unit. By solving $\Delta=0$, we can obtain the radii of the black hole horizons
\begin{equation}
r_\pm=G_{\rm N}(1+\alpha)M\biggl[1\pm\sqrt{1-\frac{a^2}{G_{\rm N}^2(1+\alpha)^2M^2}-\frac{\alpha}{1+\alpha}}\biggr].
\end{equation}
For simplicity, we will set $G_{\rm N}=1$ throughout this paper.

\section{Photon orbits}
\label{phorb}

In this section, we would like to give a brief introduction to the photon orbits in the background of a Kerr-MOG black hole.

\subsection{Equations of geodesic motion}

The motion of a photon moving in the black hole background is described by the geodesic equation
\begin{equation}
\label{photoneqmotion}
\frac{d^2x^\mu}{d\lambda^2}+{\Gamma^\mu}_{\alpha\beta}\frac{dx^\alpha}{d\lambda}\frac{dx^\beta}{d\lambda}=0,
\end{equation}
where $\lambda$ and $\Gamma^{\mu}_{\;\;\alpha\beta}$ are the affine parameter and the Christoffel symbols of the background geometry. Combined with initial conditions, the geodesic equation (\ref{photoneqmotion}) can be solved for the specified metric. However, this approach is very difficult to deal with. Nevertheless, we can adopt the Hamilton-Jacobi approach to solve the geodesics.

The Hamilton-Jacobi equation in this spacetime with the metric tensor $g^{\mu\nu}$ is given by
\begin{equation}
\frac{\partial S}{\partial\lambda}=-\frac{1}{2}g^{\mu\nu}\frac{\partial S}{\partial x^\mu}\frac{\partial S}{\partial x^\nu}.\label{HJ}
\end{equation}
Considering that there are two Killing fields $\xi_{t, \phi}=\partial_{t, \phi}$, the Jacobi action $S$ for the photons has the following form
\begin{equation}
S=-Et+J_{\rm z}\phi+S_{r}(r)+S_\vartheta(\vartheta),\label{actionn}
\end{equation}
where $E$ and $J_{\rm z}$ are, respectively, the energy and the angular momentum in the direction of the axis of symmetry. Substituting (\ref{actionn}) into (\ref{HJ}), one can obtain the equations of motion \cite{Moffat3}
\begin{eqnarray}
\rho^{2}\frac{dt}{d\lambda}&=&a(J_{\rm z}-aE\sin^2\vartheta)+\frac{r^2+a^2}{\Delta}[(r^2+a^2)E-aJ_{\rm z}],\\
\rho^{2}\frac{d\phi}{d\lambda}&=&\frac{J_{\rm z}}{\sin^2\vartheta}-aE+\frac{a}{\Delta}[(R^2+a^2)E-aJ_{\rm z}],\\
\label{Requation}
\rho^{2}\frac{dr}{d\lambda}&=&\sqrt{{\cal R}},\\
\rho^{2}\frac{d\theta}{d\lambda}&=&\sqrt{\Theta}.
\end{eqnarray}
Here, ${\cal R}$ and $\Theta$ are given by
\begin{eqnarray}
{\cal R}&=&[(r^2+a^2)E-aJ_{\rm z}]^2-\Delta[{\cal K}+(J_{\rm z}-aE)^2],\\
\quad \Theta&=&{\cal K}+\cos^2\vartheta\biggl(a^2E^2-\frac{J_{\rm z}^2}{\sin^2\vartheta}\biggr),
\end{eqnarray}
with ${\cal K}$ denoting the Carter constant. Since the spacetime is asymptotically flat, the photon path is a straight line at infinity. However, when there a black hole is placed between the observer and the light source, the light will reach the observer after deflecting due to the strong gravitational field of the black hole. And we are going to study this case in the next part.

\subsection{Circular photon orbits}

In this subsection, we will discuss the radial motion of photon. The motion of photon is determined by two impact parameters $\xi=J_{\rm z}/E$ and $\eta={\cal K}/E^2$. From Eq. (\ref{Requation}), we can obtain the circular photon orbits, which are very useful on determining the shape of the black hole shadow. The conditions for these orbits are
\begin{equation}
 {\cal R}(r)=0, \quad \frac{d\cal R}{dr}=0.
\end{equation}
Solving them, we get
\begin{eqnarray}
\xi(r)&=&\frac{M(1+\alpha)[a^2+r(M\alpha-r)]+r\Delta}{a(M-r+M\alpha)},\label{a1}\\
\eta(r)&=&\frac{r^2}{a^2(M-r+M\alpha)^2}\bigg[4Mr(1+\alpha)\big(a^2+M^2\alpha(1+\alpha) \big)-4M^2\alpha(1+\alpha)\Delta-r^2\big(r-3M(1+\alpha)\big)^2\bigg].\label{a2}
\end{eqnarray}
These two impact parameters are the key to determine the boundary of the black hole shadow, and we will use them later.

\section{Shadows of MOG black hole}
\label{sshadow}

As we mentioned above, when there is a black hole located between a light source and an observer, the light will be deflected due to the strong gravitational field of the black hole. In general, there are different cases when a photon emitted by a light source passes a black hole from infinity. For the first case, the photon with large orbital angular momentum will turn back at some turning points and reach the observer located at infinity. For the second case, the photon with small orbital angular momentum will fall into the black hole. Thus the light source is invisible to the observer. If the light sources are located at infinity and distributed uniformly in all directions, the observer will see a dark zone, known as the black hole shadow corresponding to the second case. In particular, among these two cases, there exists a critical case where the photon will move around the black hole all the time. And this critical case just relates with the boundary of the dark zone. Since the information of the shadow is encoded in its boundary, we mainly focus on the critical case.

\subsection{Apparent shape}

\begin{figure*}
\begin{center}
\subfigure[]{\label{circle1}
\includegraphics[width=6cm,height=6cm]{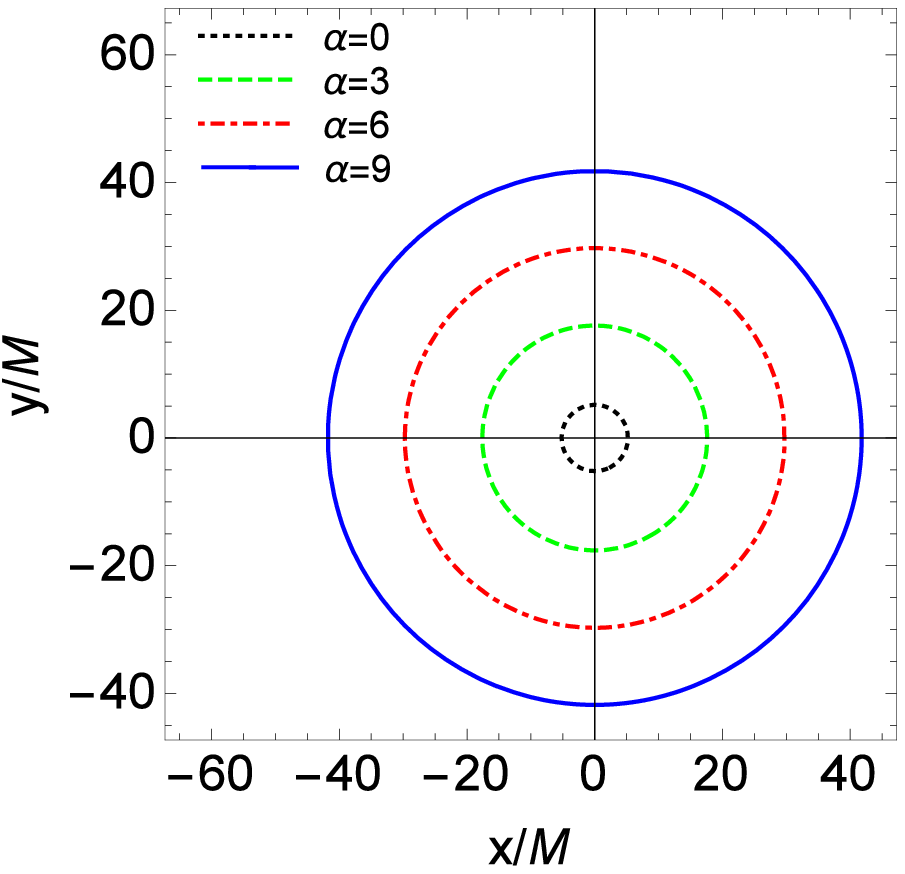}}
\subfigure[]{\label{circle2}
\includegraphics[width=6cm,height=6cm]{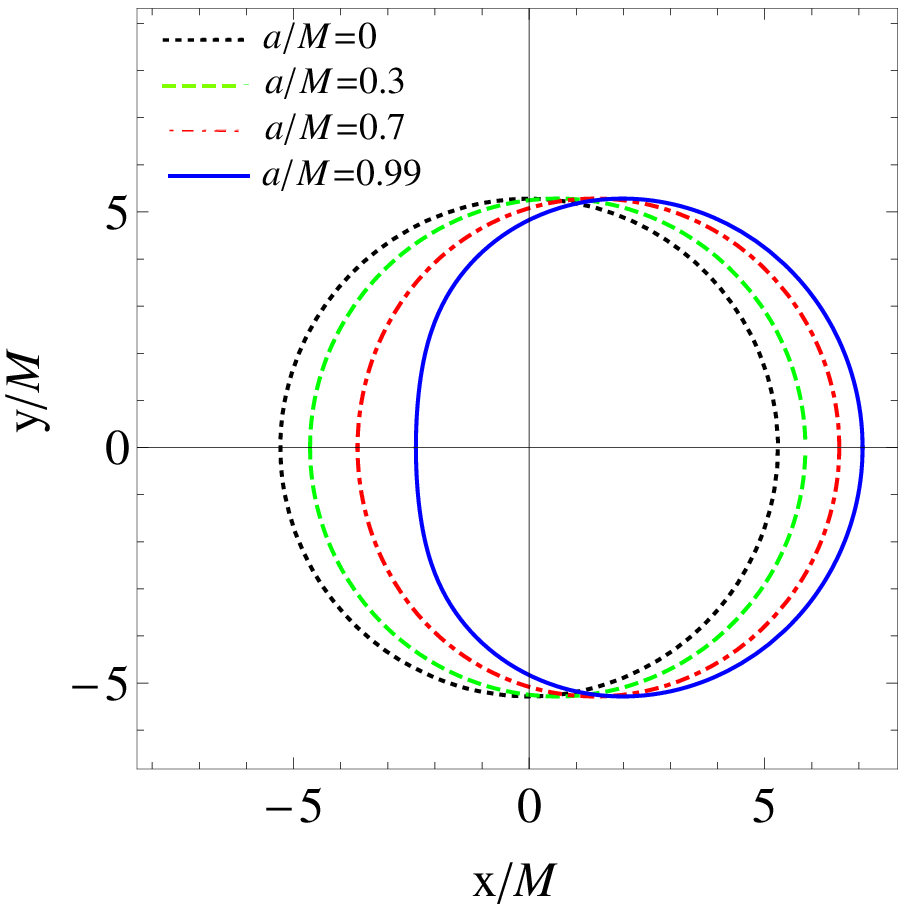}}
\end{center}
\caption{Shadows cast by black holes. (a) The non-rotating case $a/M=0$. (b) $\alpha=0.02$, $\theta=\frac{\pi}{2}$.}
\end{figure*}

The border of the shadow defines the apparent shape of the black hole. To study the shadows, we adopt the celestial coordinates:
\begin{eqnarray}
x&=&-\xi\csc\theta,\\
y&=&\pm\sqrt{\eta+a^2\cos^2\theta-\xi^2\cot^2\theta},
\end{eqnarray}
where $\theta$ is the angle between the rotation axis of the black hole and the line of sight of the observer. In the special case where the observer is on the equatorial plane of the black hole with the inclination angle $\theta=\frac{\pi}{2}$, one can easily get
\begin{equation}
 x=-\xi,\quad
 y=\pm\sqrt{\eta}.
\end{equation}
For the non-rotating black hole case of $a/M=0$, the parameter $r$ in Eqs. (\ref{a1}) and (\ref{a2}) will take the radius of the photon sphere
\begin{equation}
 r_{\texttt{ps}}=\frac{3}{2}(1+\alpha)M\left(1+\sqrt{1-\frac{8\alpha}{9(1+\alpha)}}\right).\label{photon}
\end{equation}
Then we have
\begin{equation}
 x^{2}+y^{2}=\frac{r_{\texttt{ps}}^4}{r_{\texttt{ps}}^2-2Mr_{\texttt{ps}}(1+\alpha)+M^2\alpha(1+\alpha)}.
 \end{equation}
From above, one can easily find that the shape of the non-rotating black hole is a standard circle, while its radius is closely dependent on the parameter $\alpha$. The shapes are plotted in Fig. \ref{circle1}. From it, we can see that the size of the shadow increases with $\alpha$. Thus comparing with the Schwarzschild black hole case, the non-rotating MOG black hole always has a larger shadow.

For the rotating black hole case of $a/M\neq0$, the shape of the shadow will be elongated in the direction of the rotation axis due to the dragging effect, so the shape will deviate from the standard circle. For example, we set $\alpha=0.02$ and vary the black hole spin parameter $a/M$ from 0 to 0.99 shown in Fig. \ref{circle2}. For a low spin, the shape is almost a circle, while the deviation becomes significant when the spin parameter approaches a large value, i.e., $a/M=0.99$. Our result shows that such pattern is the same as that for the Kerr black hole case \cite{Chandrasekhar,Hioki2}.

\begin{figure*}
\begin{center}
\subfigure[]{
\includegraphics[width=6cm]{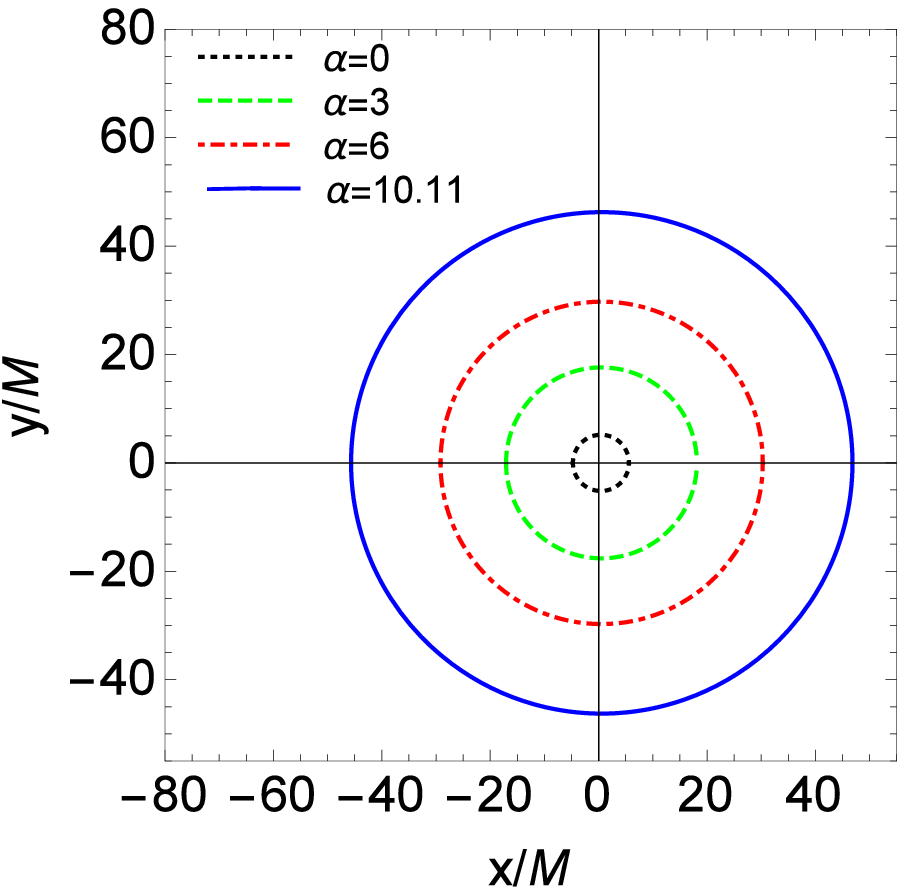}}
\subfigure[]{
\includegraphics[width=6cm]{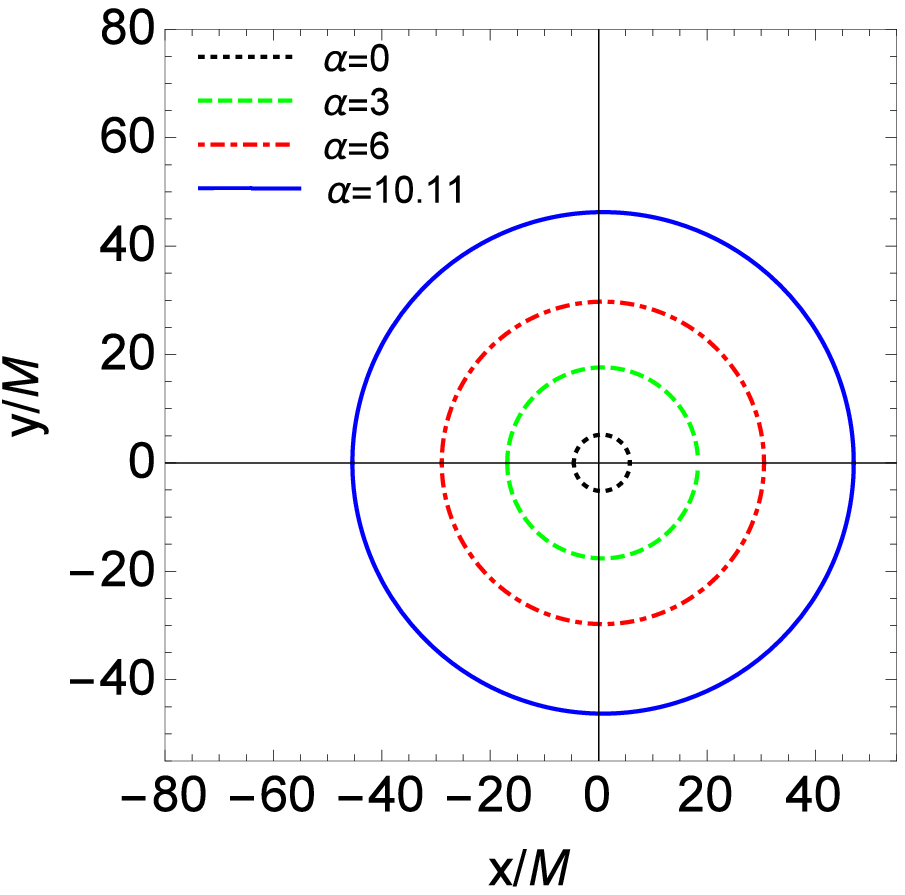}}\\
\subfigure[]{
\includegraphics[width=6cm]{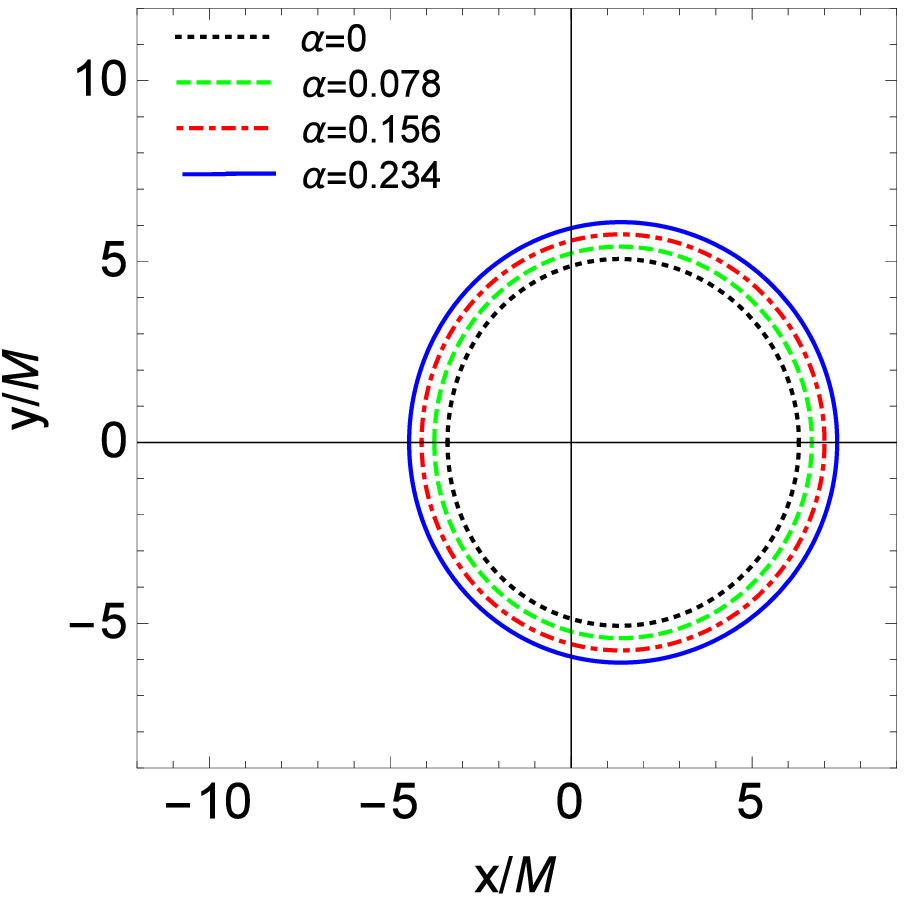}}
\subfigure[]{
\includegraphics[width=6cm]{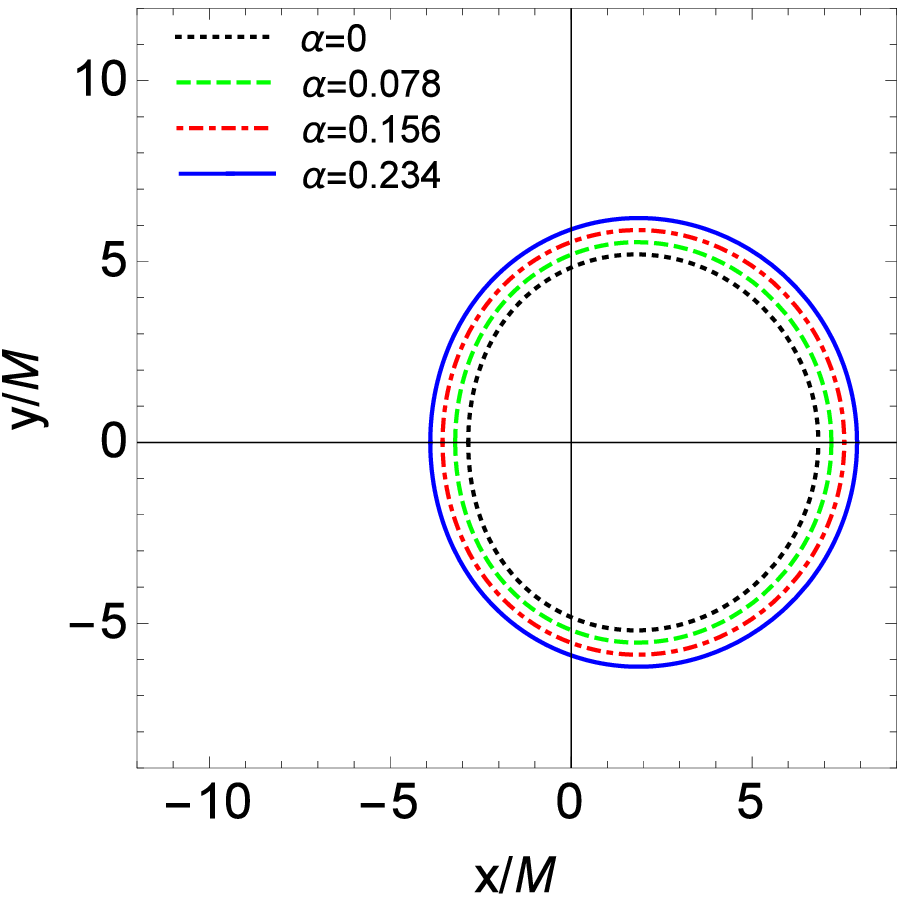}}\\
\subfigure[]{
\includegraphics[width=6cm]{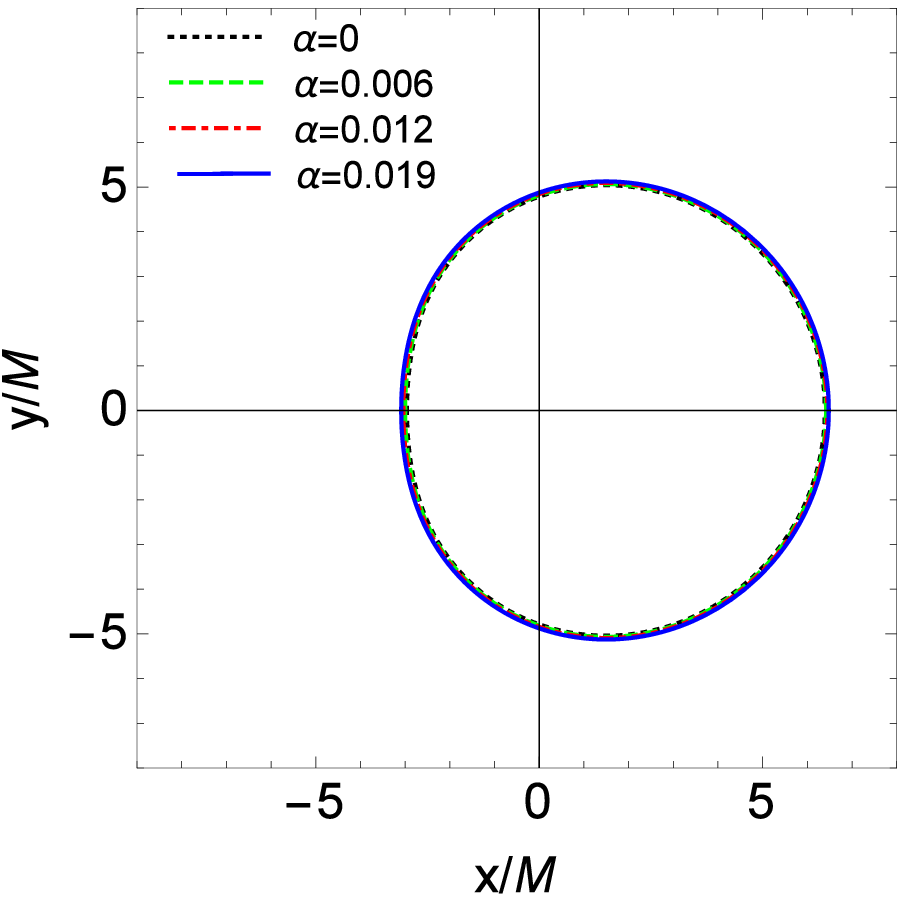}}
\subfigure[]{
\includegraphics[width=6cm]{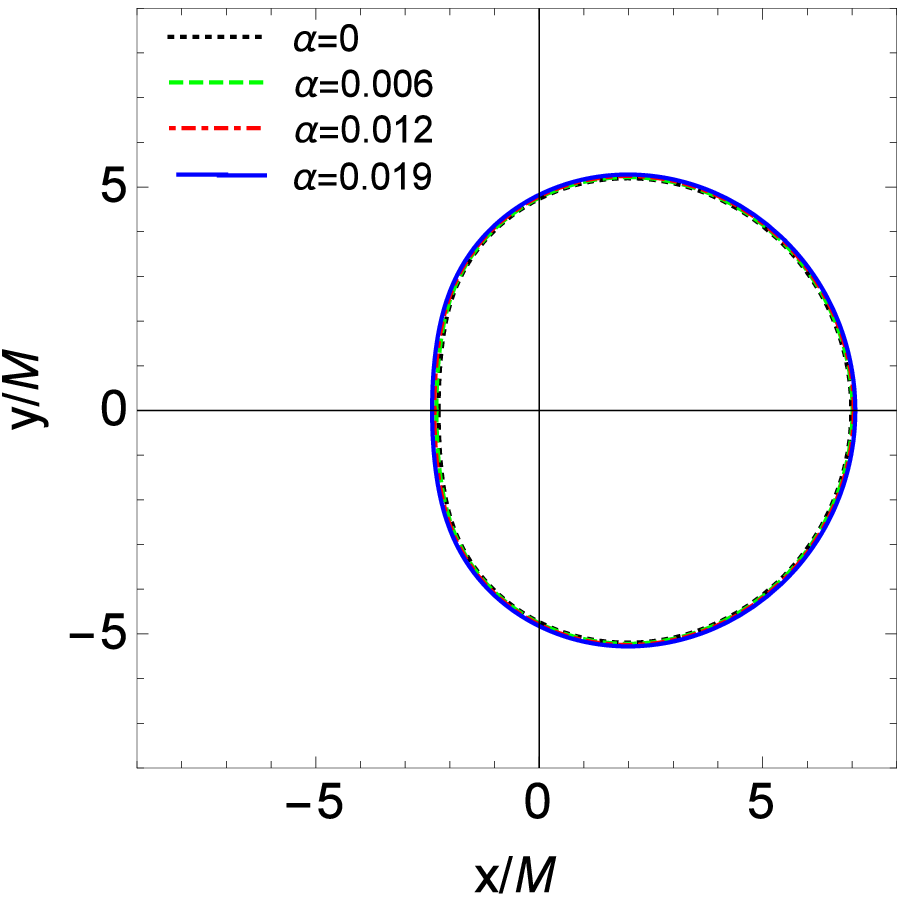}}
\caption{The shapes of the black hole shadow for different values of the parameters. (a) $a/M=0.3$,
$\theta=\frac{\pi}{4}$. (b) $a/M=0.3$,
$\theta=\frac{\pi}{2}$. (c) $a/M=0.9$,
$\theta=\frac{\pi}{4}$. (d) $a/M=0.9$,
$\theta=\frac{\pi}{2}$. (e) $a/M=0.99$,
$\theta=\frac{\pi}{4}$. (f) $a/M=0.99$,
$\theta=\frac{\pi}{2}$.}\label{circle3}
\end{center}
\end{figure*}

For clarity, we plot the shapes of the black hole shadow in Fig. \ref{circle3} for different values of the spin parameter $a/M$ and the inclination angle $\theta$. All the figures confirm that the black hole shadow closely depends on both $a/M$ and $\theta$. Moreover, the more the spin parameter $a/M$ of the black hole approaches its maximum, the more the deformed of the shape.

\subsection{Observables}

In order to determine or test the parameters of the astronomical black hole by using the shadow, constructing astronomical observables is necessary. One of the properties of the observables must be easily measured and reliable. There are different observables constructed by different groups \cite{Hioki2,Johannsen,Abdujabbarov,Bambi}. For convenience, we follow Hioki and Maeda \cite{Hioki2}, where two observables, the radius $R_{\rm s}$ and the distortion parameter $\delta_{\rm s}$, are introduced. The quantity $R_{\rm s}$ is the radius of a reference circle passing by three points: the top point $(x_{\rm t}, y_{\rm t})$, the bottom point $(x_{\rm b},y_{\rm b})$, and the right point $(x_{\rm r},0)$ of the shadow, which corresponds to the unstable retrograde circular orbit seen by an observer located on the equatorial plane. Another quantity is the distortion parameter $\delta_{\rm s}=D/R_{\rm s}$, where $D$ is the difference between the
endpoints of the reference circle and of the shadow, and both of them are at the opposite side of the point $(x_{\rm r},0)$. The radius $R_{\rm s}$ basically gives the approximate size of the shadow, and $\delta_{\rm s}$ measures its deformation with respect to the reference circle. If the inclination angle $\theta$ is independently known, precise enough measurements of $R_{\rm s}$ and $\delta _{\rm s}$ could serve, in principle, to obtain the spin parameter $a$ and the parameter $\alpha$. Moreover, in order to avoid the degenerate of the shadow caused by the black hole parameters, another distortion parameter $\varepsilon$ is needed \cite{Tsukamoto}, which measures the distortion on the left side of the black hole shadow formed by the unstable retrograde circular orbit. We denote the corresponding point as $(x_{\rm h}, y_{\rm h})$, which is just the point that the horizontal line of $y_{\rm h}=y_{\rm t}/2$ cuts the shadow at the opposite side of $(x_{\rm r}, 0)$. Combined with these three observables, it is expected that the characterizations of the MOG black hole and the inclination angle $\theta$ of the observer can be well determined.

Taking a simple algebra calculation, one can find that the center ($x_{\rm c}$, $y_{\rm c}$) of the reference circle is at \begin{eqnarray}
 x_{\rm c}&=&\frac{x_{\rm r}^{2}-x_{\rm t}^{2}-y_{\rm t}^{2}}{2(x_{\rm r}-x_{\rm t})},\\
 y_{\rm c}&=&0.
\end{eqnarray}
Note that $y_{\rm c}=0$ is the result of the symmetry of the shadow. Further, these three observables can be expressed as
\begin{eqnarray}
 R_{\rm s}&=&\frac{(x_{\rm t}-x_{\rm r})^{2}+y_{\rm t}^{2}}{2(x_{\rm r}-x_{\rm t})},\label{rrss}\\
 \delta_{\rm s}&=&\frac{(x_{\rm l}-\tilde{x}_{\rm r})}{R_{\rm s}},\\
 \epsilon&=&1-\frac{\sqrt{(x_{\rm h}-x_{\rm c})^{2}+y_{\rm t}^{2}/4}}{R_{\rm s}},
\end{eqnarray}
where $(\tilde{x}_{\rm r}, 0)$ and $(x_{\rm l}, 0)$ are the points
where the reference circle and the contour of the shadow cut the horizontal
axis at the opposite side of $(x_{\rm r}, 0)$, respectively. For the shadow cast by the non-rotating black hole, the contour of the shadow will meet the reference circle, and thus there will be no distortion, i.e., $\epsilon=\delta_{\rm s}=0$.

\begin{figure*}
\begin{center}
\subfigure[]{
\includegraphics[width=6cm,height=6cm]{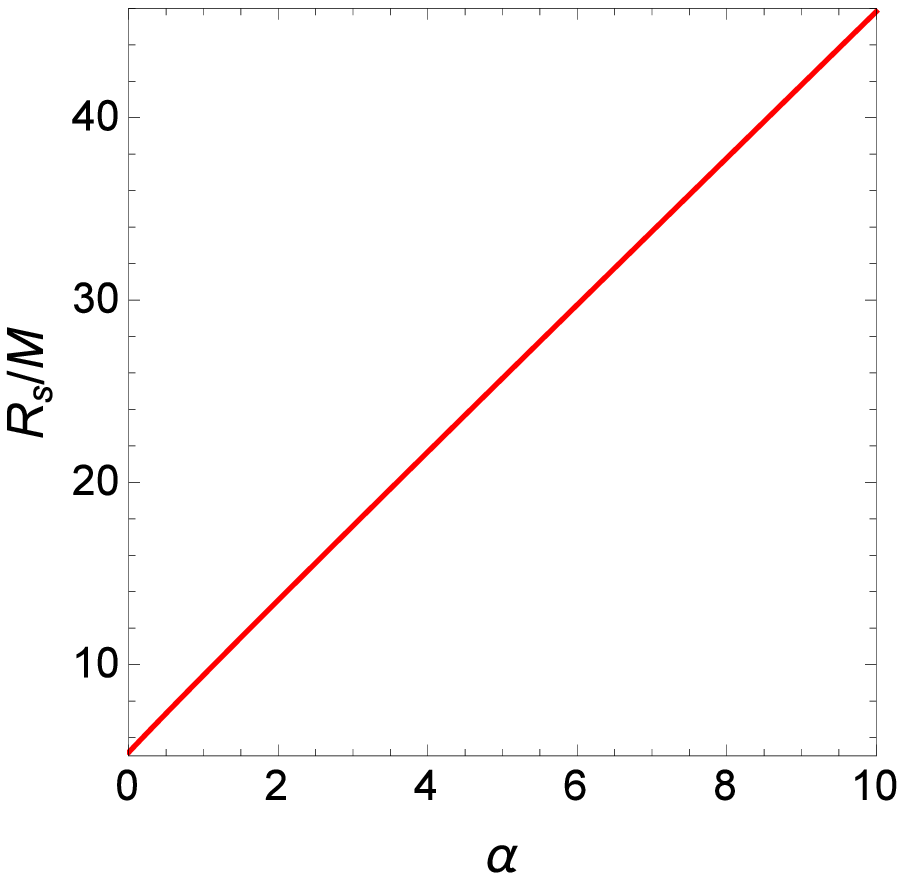}}
\subfigure[]{
\includegraphics[width=6cm,height=6cm]{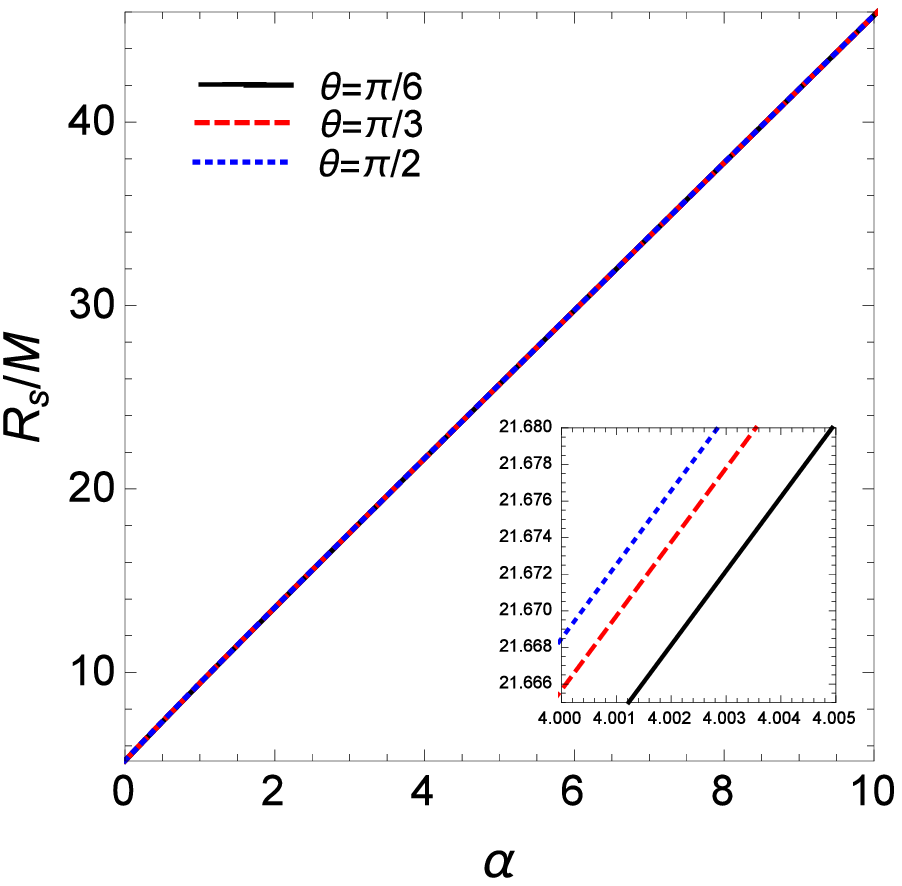}}\\
\subfigure[]{
\includegraphics[width=6cm,height=6cm]{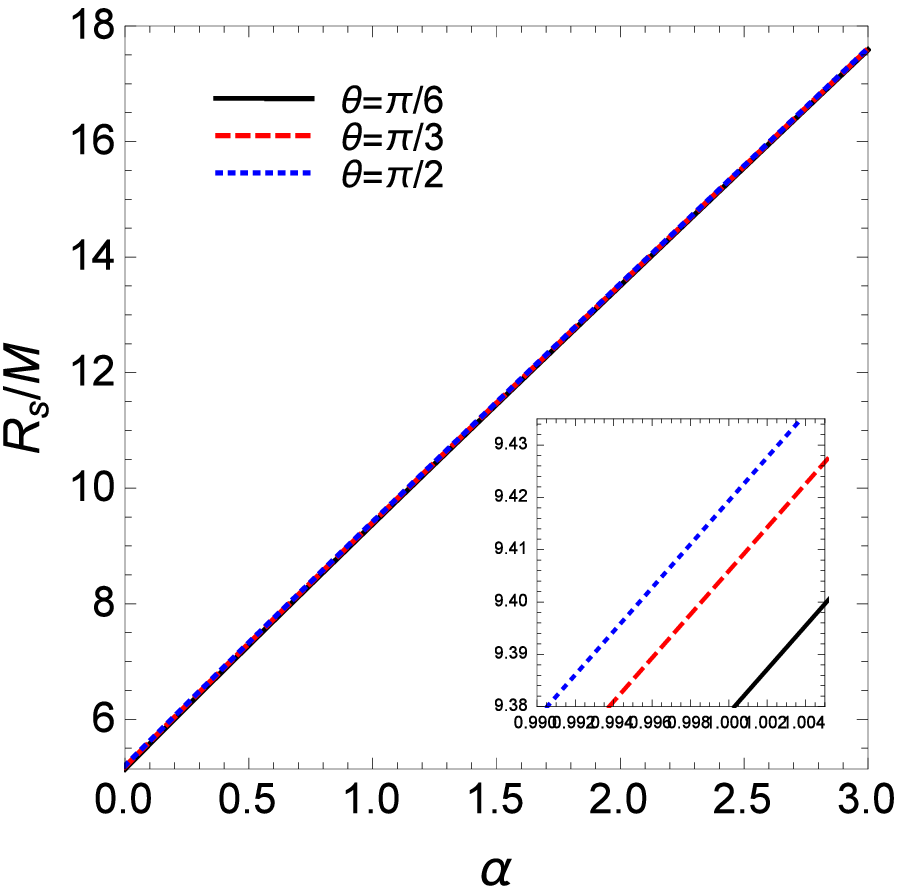}}
\subfigure[]{
\includegraphics[width=6cm,height=6cm]{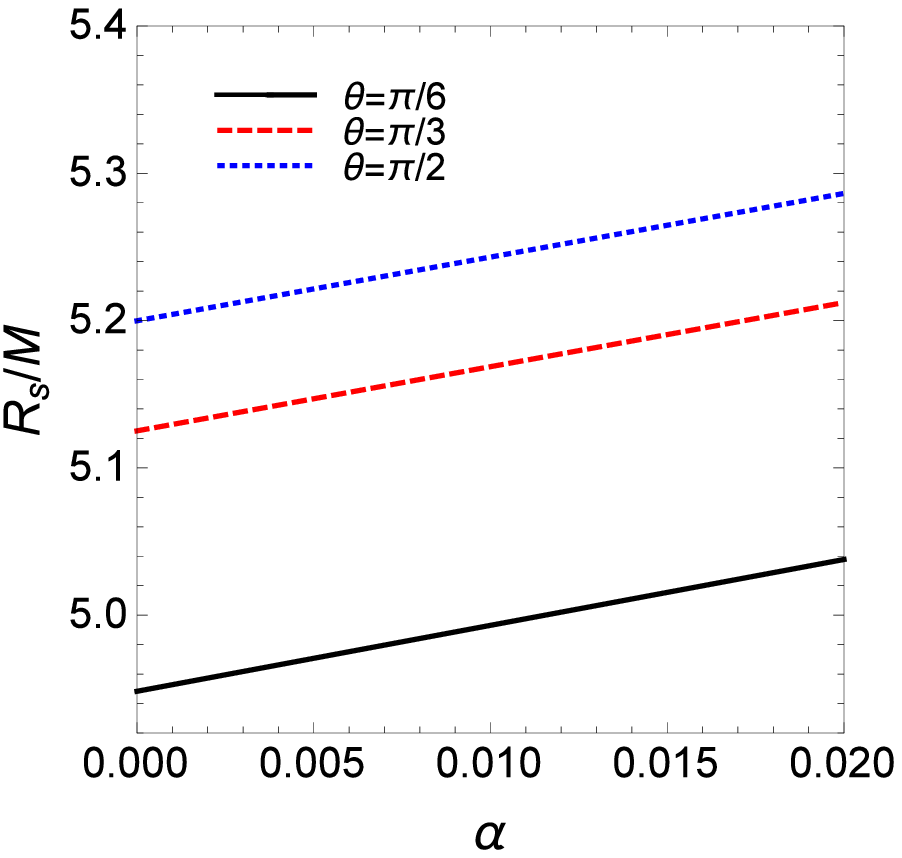}}
\caption{The radius $R_{\rm s}$ of the black hole shadow against the parameter $\alpha$ for different values of the spin parameter $a/M$, and the inclination angle $\theta$. The black solid lines, red dashed lines, and blue dotted lines are for $\theta=\frac{\pi}{6}$, $\frac{\pi}{3}$, and $\frac{\pi}{2}$, respectively. (a) $a/M=0$. (b) $a/M=0.3$. (c) $a/M=0.5$. (d) $a/M=0.99$.}\label{Rs}
\end{center}
\end{figure*}

\begin{figure}
  \begin{center}
\subfigure[]{\label{radiusps1}
\includegraphics[width=6cm,height=6cm]{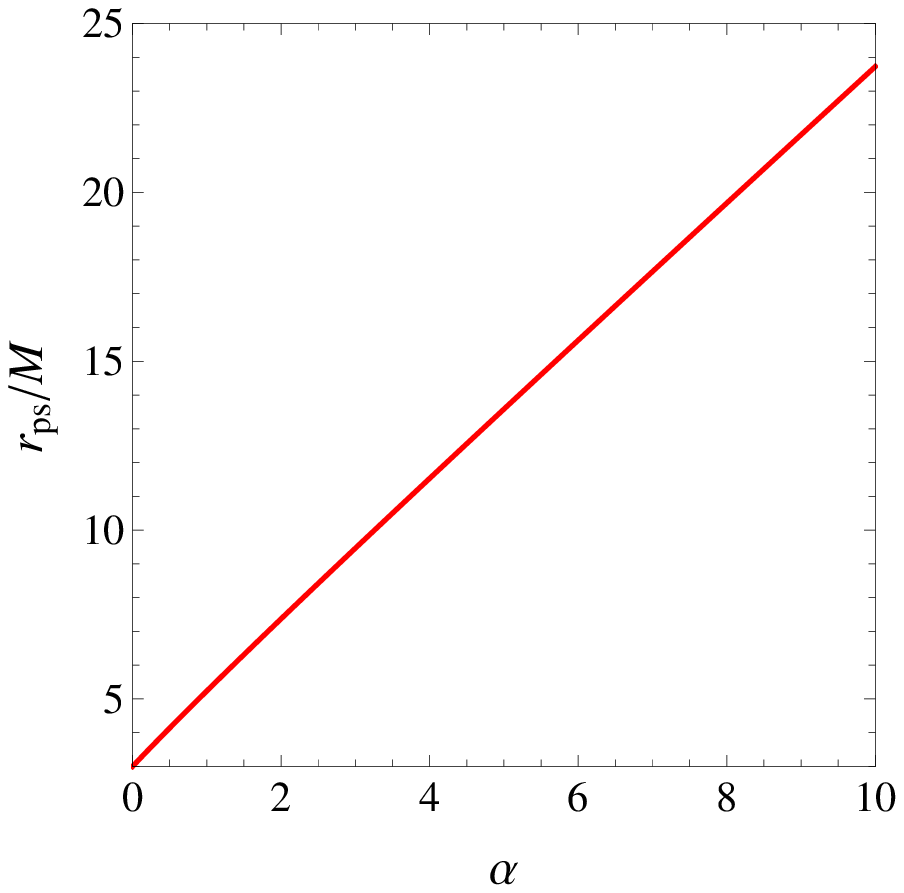}}
\subfigure[]{\label{radiusps2}
\includegraphics[width=6cm,height=6cm]{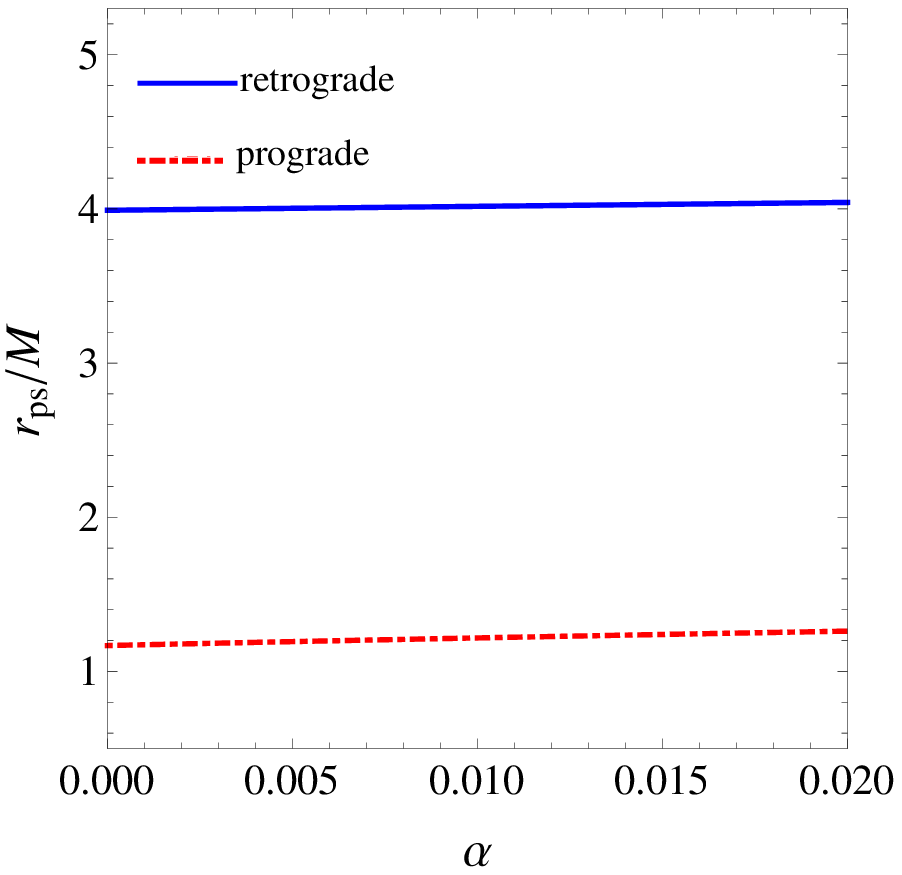}}
  \caption{The radius of the photon sphere $r_{\rm ps}$ against $\alpha$. (a) $a/M=0$. (b) $a/M=0.99$.}
  \end{center}
\end{figure}

\begin{figure*}
\begin{center}
\subfigure[]{
\includegraphics[width=6cm,height=6cm]{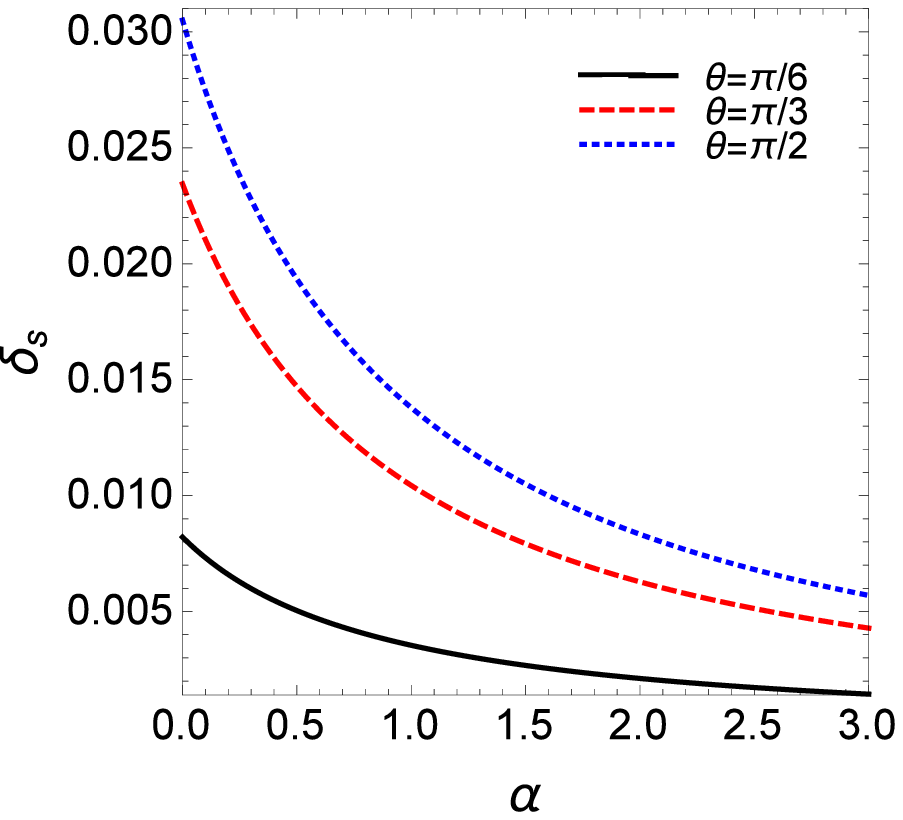}}
\subfigure[]{
\includegraphics[width=6cm,height=6cm]{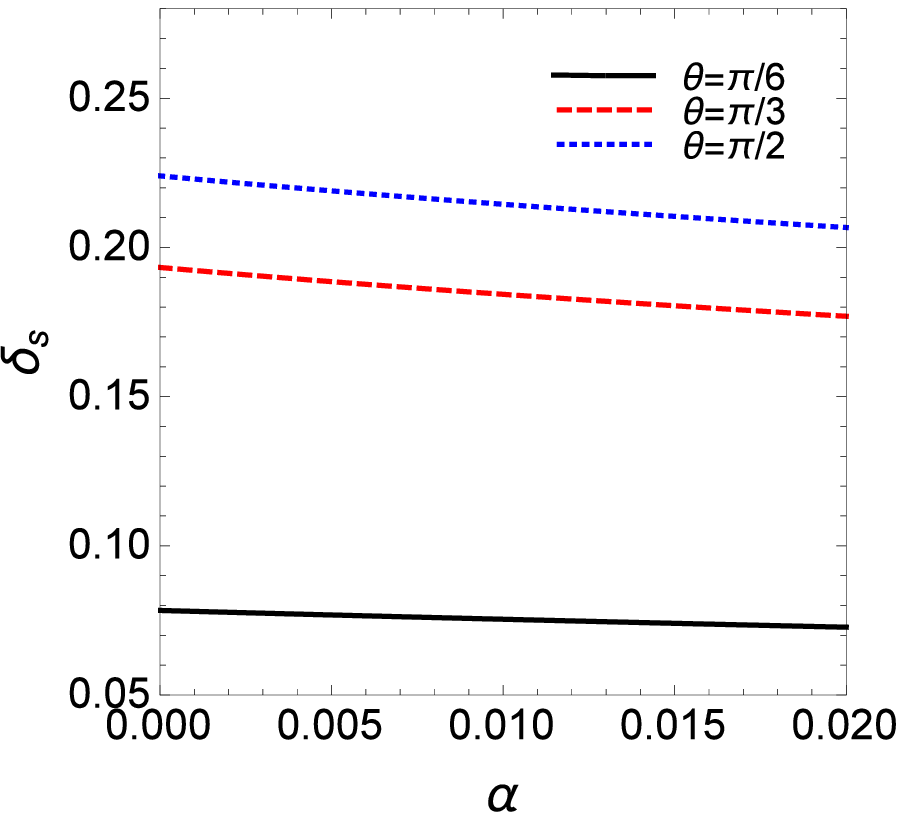}}
\caption{The distortion $\delta_{\rm s}$ of the black hole shadow against $\alpha$ for different values of the spin parameter $a/M$, and the inclination angle $\theta$. The black solid lines, red dashed lines, and blue dotted lines are for $\theta=\frac{\pi}{6}$, $\frac{\pi}{3}$, and $\frac{\pi}{2}$, respectively. (a) $a/M=0.5$. (b) $a/M=0.99$.}\label{delta}
\end{center}
\end{figure*}

\begin{figure*}
\begin{center}
\subfigure[]{
\includegraphics[width=6cm,height=6cm]{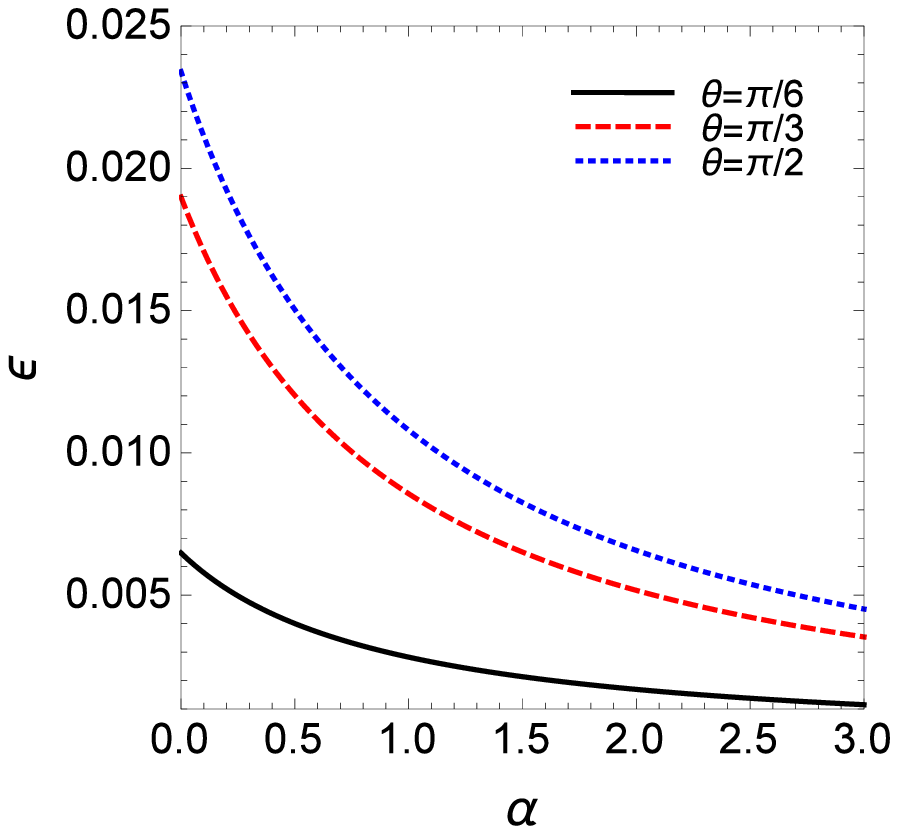}}
\subfigure[]{
\includegraphics[width=6cm,height=6cm]{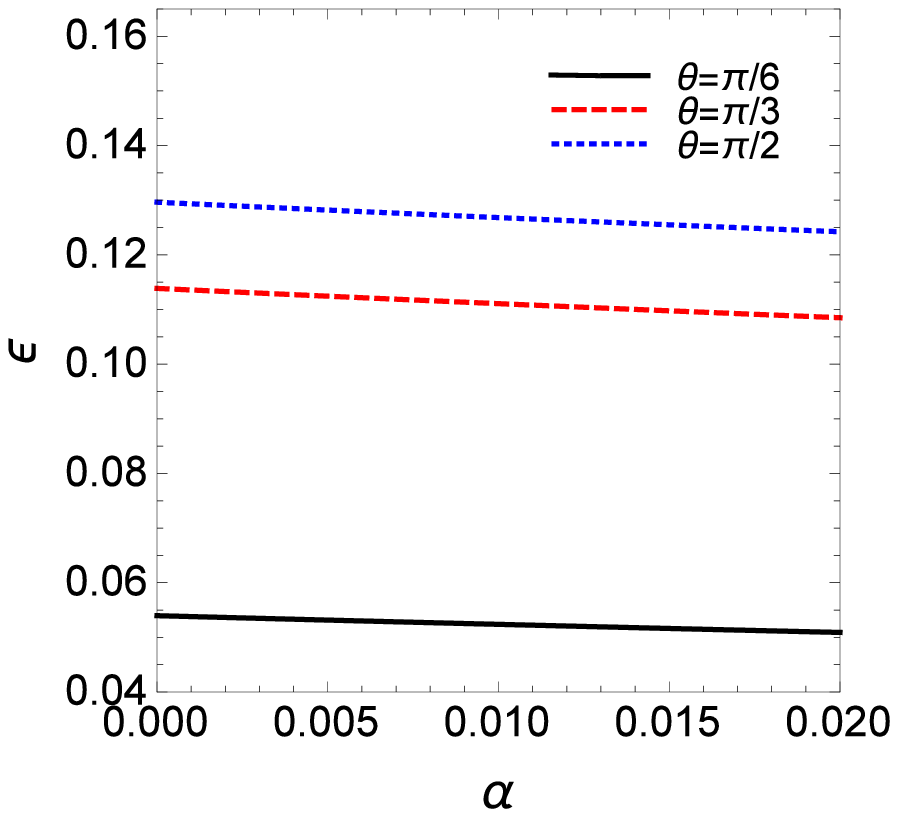}}
 \caption{The $\varepsilon$ of the black hole shadow against $\alpha$ for different values of the spin parameter $a/M$, and the inclination angle $\theta$. The black solid lines, red dashed lines, and blue dotted lines are for $\theta=\frac{\pi}{6}$, $\frac{\pi}{3}$, and $\frac{\pi}{2}$, respectively. (a) $a/M=0.5$. (b) $a/M=0.99$.}\label{e}
\end{center}
\end{figure*}

We numerically calculate these three observables, and the results are presented in Figs. \ref{Rs}, \ref{delta} and \ref{e}. The cases for the inclination angle $\theta=\frac{\pi}{6}$, $\frac{\pi}{3}$, and $\frac{\pi}{2}$ are, respectively, described by the black solid lines, red dashed lines, and blue dotted lines. The radius $R_{\rm s}$ of the shadow is clearly shown in Fig. \ref{Rs}. From it, we can find that for fixed inclination angle $\theta$ and spin parameter $a/M$, $R_{\rm s}$ increases with the parameter $\alpha$. This is because that the gravity of the black hole increases with $\alpha$. So the MOG black hole always has a larger shadow than the GR black hole. Interestingly, we can also find that the radius $R_{\rm s}$ increases with $\theta$, which means that the observer located on the equatorial plane will get the biggest shadow. On the other hand, we also study the radius of the photon sphere $r_{\rm ps}$. When $a/M=0$, the radius $r_{\rm ps}$ of the photon sphere is given in Eq. (\ref{photon}), and its behavior is plotted in Fig. \ref{radiusps1} as a function of the parameter $\alpha$. From the figure, one can find that when $r_{\rm ps}$ increases with $\alpha$, which indicates that the gravity increases with $\alpha$. This is because that the larger $\alpha$ is, the larger the effective gravitation constant $G$.

The two distortion parameters $\delta_{\rm s}$ and $\epsilon$ are clearly displayed in Figs. \ref{delta} and \ref{e}. As expected, $\delta_{\rm s}$ and $\epsilon$ are nonzero for rotating black holes. Obviously, $\delta_{\rm s}$ and $\epsilon$ decrease with the parameter $\alpha$. For fixed $a/M=0.5$ and $\theta=\pi/2$, $\delta_{\rm s}$ approaches to 13\% and $\epsilon$ tends to 11\% for $\alpha=1$. So comparing with the Kerr black hole in GR, the rotating MOG black hole gets less deformed. This can provide us a possible way to distinguish these two black holes. The decrease of the distortion with $\alpha$ is mainly caused by the radii $r_{\rm ps}$ of the retrograde light ring and the prograde light ring. Taking $a/M=0.99$ as an example, we plot $r_{\rm ps}$ for retrograde light ring and the prograde light ring in Fig. \ref{radiusps2}. It is obvious that, for both the retrograde and prograde cases, $r_{\rm ps}$ increases with $\alpha$. However, the difference is that the slope is about 4.68 for the prograde case, and 2.55 for the retrograde case, which means that the radii of the two light rings vary at different speeds of $\alpha$. The prograde light ring has a faster growth rate than that of the retrograde one, and which will lead to that, with the increase of $\alpha$, the horizontal diameter of the shadow gets a little longer. Then, the distortions of the shadows decrease.

\begin{figure*}
\begin{center}
\subfigure[]{
\includegraphics[width=6cm,height=6cm]{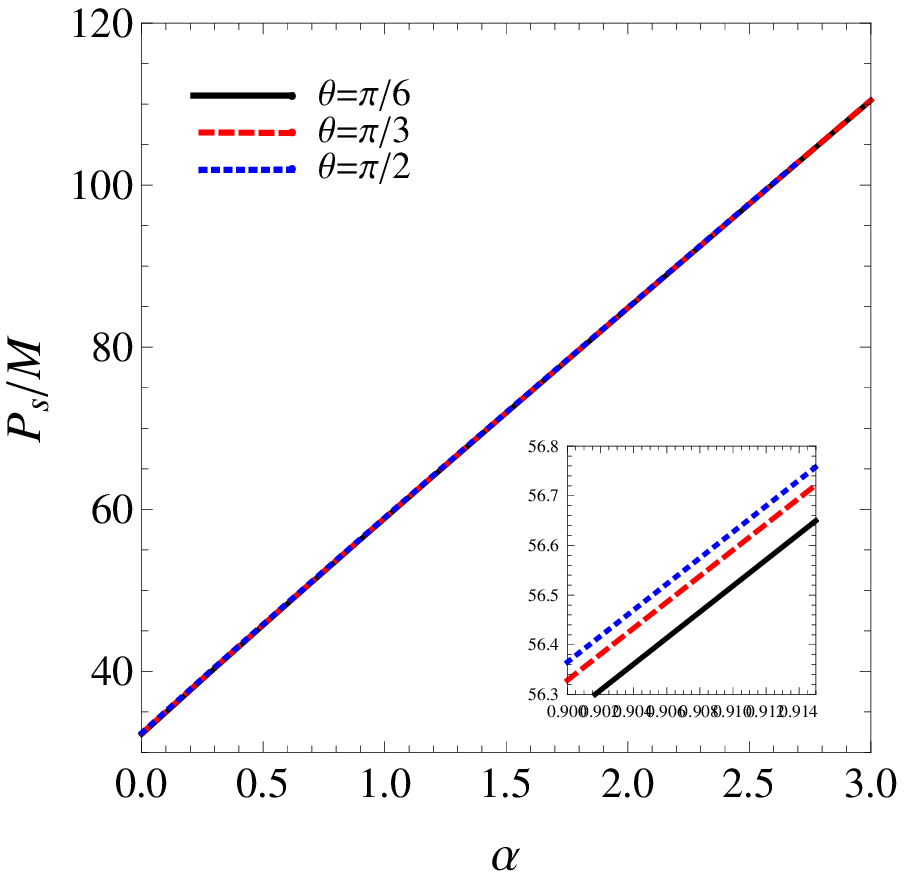}}
\subfigure[]{
\includegraphics[width=6cm,height=6cm]{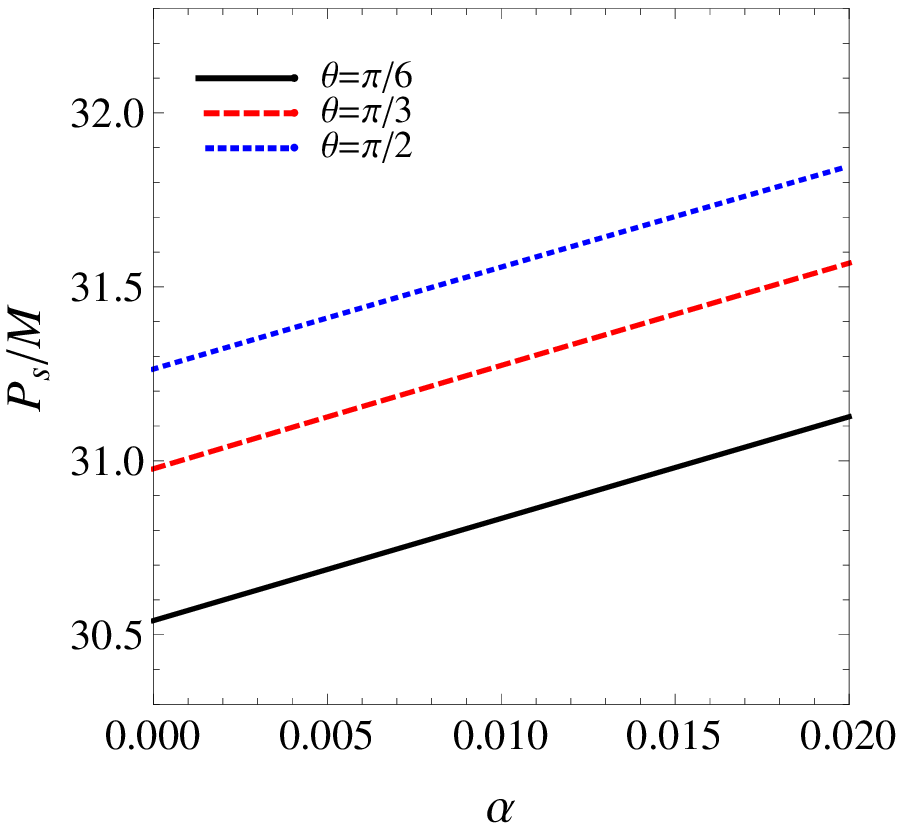}}
\caption{The perimeter $P_{\rm s}$ of the black hole shadow against $\alpha$ for different values of the spin parameter $a/M$, and the inclination angle $\theta$. The black solid lines, red dashed lines, and blue dotted lines are for $\theta=\frac{\pi}{6}$, $\frac{\pi}{3}$, and $\frac{\pi}{2}$, respectively. (a) $a/M=0.5$. (b) $a/M=0.99$.}\label{perimeter}
\end{center}
\end{figure*}

From above, we can find that there are four parameters, the mass $M$, spin $a$, coupling parameter $\alpha$, and viewing angle $\theta$, to be determined. In general, the black hole mass can be well measured by the motion of the star around it. On the other side, the black hole mass is closely related with the size of the shadow. For example, for the Schwarzschild black hole, the perimeter $P_{\rm s}$ of the shadow is
\begin{eqnarray}
 P_{\rm s}=6\pi\sqrt{3} M,\label{scs}
\end{eqnarray}
which implies that the larger the black hole, the longer the perimeter of its shadow. Thus the perimeter $P_{\rm s}$ can act as an observable to measure the black hole parameter. In the following, we will study $P_{\rm s}$ for the shadow. For a certain shadow, its perimeter can be calculated with \cite{Abdujabbarov,Mann3}
\begin{equation}
P_{\rm s}=2\int^{r_2}_{r_1}\sqrt{(\partial_rx)^{2}+(\partial_ry)^{2}}dr.
\end{equation}
Here $r_1$ and $r_2$ denote the radii of light rings for prograde and retrograde photons, respectively. For the non-rotating case $a/M=0$, the perimeter has an analytical form
\begin{equation}
P_{\rm s}=\frac{\pi  \sqrt{\alpha +1} \left(3
   \sqrt{\alpha +1}+\sqrt{\alpha +9}\right)^2}{\sqrt{2} \sqrt{\alpha +\sqrt{(\alpha +1)
   (\alpha +9)}+3}}M.
\end{equation}
When $\alpha=0$, it reduces to the Schwarzschild black hole case (\ref{scs}). A detailed study shows that $P_{\rm s}$ increases with $\alpha$. Moreover, for fixed $\alpha$, $P_{\rm s}\propto M$. Thus $P_{\rm s}$ can be used to test the black hole parameter. For the rotating case, we can obtain the numerical values of the perimeter. For $a/M=0.5$ and $a/M=0.99$, we show the results in Fig. \ref{perimeter}. From it, we find that $P_{\rm s}$ increases with the parameter $\alpha$ and the viewing angle $\theta$.

\begin{figure*}
\begin{center}
\subfigure[]{
\includegraphics[width=8cm]{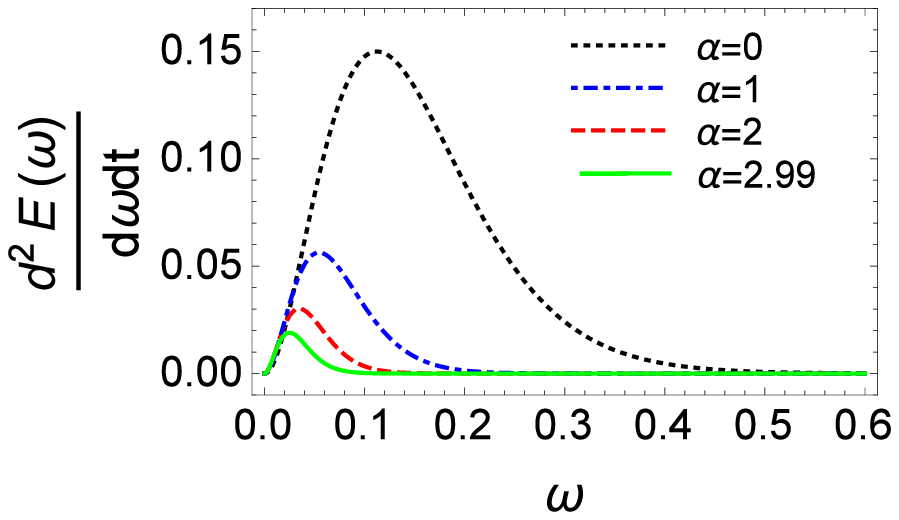}}
\subfigure[]{
\includegraphics[width=8cm]{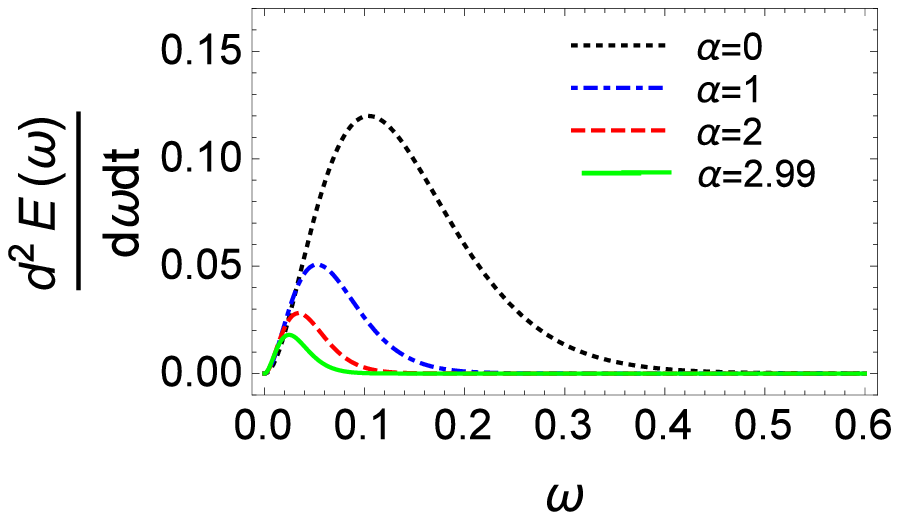}}
\caption{Behaviors of the energy emission rate $\frac{d^2E(\omega)}{d\omega dt}$ for $\alpha=0, 1, 2, 2.99$. (a) The non-rotating case $a/M=0$. (b) The rotating case $a/M=0.5$.}\label{T}
\end{center}
\end{figure*}

In Ref. \cite{Wei}, we proposed that the area of the black hole shadow observed by the observer on the equatorial plane located at infinity approximately equals to the high energy absorption cross section, and this assumption has been carried out for the Einstein-Maxwell-Dilaton-Axion black hole. Adopting the assumption, in high energy case, the energy emission rate of the black hole is
\begin{equation}
\frac{d^2E(\omega)}{d\omega dt}=\frac{2\pi^3 R_{\rm s}^2}{e^{\omega/T_{H}}-1}\omega^3,
\end{equation}
where $R_{\rm s}$ is given in Eq. (\ref{rrss}). The Hawking temperature $T_{H}$ can be calculated by
\begin{eqnarray}
T_{H}&=&\lim_{\theta=0,r\to r_+}\frac{\partial_r\sqrt{g_{tt}}}{2\pi\sqrt{g_{rr}}}\\
  &=&\frac{-2a^2+M^2(1+\alpha)(2+2X+\alpha X)}{2M^3\pi(1+\alpha)(2+\alpha+2X+2\alpha X)^2},
\end{eqnarray}
with $X=\frac{\sqrt{-a^2+M^2(1+\alpha)}}{M(1+\alpha)}$. Then, we describe the energy emission rate against the frequency $\omega$ for $a/M=0$ and $a/M=0.5$ in Fig. \ref{T}. Obviously, there exists a peak for each curve. With the increase of the parameter $\alpha$, the peak decreases and shifts to the low frequency. For fixed parameter $\alpha$, the peak decreases with spin parameter $a/M$.

 \section{Conclusions and discussions}
 \label{discussions}

We studied the black holes predicted by modified gravity theory for both rotating and non-rotating cases. The black hole has an effect on the photons moving around it due to its strong gravitational field. When photons from infinity approach a black hole, their directions will change. Those photons having large orbital angular momentum will deflect due to gravity and reach the observer, while photons with small orbital angular momentum will be captured by the black hole, and form a dark zone in the sky. The shadow is the optical appearance of the black hole, and its apparent shape is considered to be the boundary of the black hole. Then one can use the formation mechanism of shadow to study the property of the black holes.

 In this paper we studied the optical features of black hole. There will be some deformations in the shape of the black hole shadow in rotating case. Both the circular shadow for the Schwarzschild-MOG black hole ($a/M=0$) and the deformed circular shadow for the Kerr-MOG black hole ($a/M\neq 0$) increase with $\alpha$. With the increase of the spin parameter $a/M$, the shadow is shifted to the right. Taking $\alpha=0.02$ and $\theta=\frac{\pi}{2}$, we clearly showed that as the value of $a/M$ increases, the size of the shadows are almost constant, but the shapes get more deformed and move to the right, see Fig. \ref{circle2}. 
 
Then we studied four observables, the radius $R_{\rm s}$, distortions $\delta_{\rm s}$, $\epsilon$, and the perimeter $P_{\rm s}$ for the shadows cast by the MOG black holes. The results show that the radius $R_{\rm s}$ and perimeter $P_{\rm s}$ of the black hole shadow increase with the viewing angle $\theta$ and the parameter $\alpha$, which indicates that the gravity of the MOG black holes increases with $\alpha$. On the other hand, the distortions $\delta_{\rm s}$ and $\epsilon$ decrease with $\alpha$ for fixed black hole spin. Thus the shadows of MOG get smaller deformed from its GR counterpart. These results will provide us a novel approach to determine the black hole parameters and to test the MOG gravity by making use of the shadow.

Moreover, due to the optical properties, the area of the black hole shadow is supposed to equal to the high-energy absorption cross section. Based on this assumption, the energy emission rate is investigated. From the figure, we found that with the increase of parameter $\alpha$, the peak decreases and shifts to the low frequency.

 It is currently believed that there are supermassive black holes in the center of many galaxies. There is now evidence that such a black hole exists in the center of the Milky Way and Sgr $A^{*}$ is a candidate. Therefore, we expect to test the nature of the supermassive black hole in our galaxy modeled with the MOG black hole solution. Especially, with the help of the Event Horizon Telescope at wavelengths around 1 mm based on VLBI \cite{Fish}, it also provides us a possible way to distinguish the MOG gravity from GR by observing the black hole shadow.

\section*{Acknowledgements}
This work was supported by the National Natural Science Foundation of China (Grants No. 11675064). We thank Prof. Yu-Xiao Liu for helpful suggestions.

\end{document}